\documentclass[prc,aps,floats,floatfix,twocolumn,showpacs,superscriptaddress,nofootinbib]{revtex4}
\usepackage{graphicx}

\newcommand{\nuc}[2]{$^{#1}${#2}}
\newcommand{\etal}{\emph{et al.}}

\begin{document}

\title{Shape coexistence in neutron-deficient Kr isotopes:
       Constraints on the single-particle spectrum
       of self-consistent mean-field models from collective excitations}

\author{M. Bender}
\affiliation{Service de Physique Nucl\'eaire Th\'eorique,
             Universit\'e Libre de Bruxelles, Case Postale 229,
             B-1050 Bruxelles, Belgium}
\affiliation{National Superconducting Cyclotron Laboratory,
             Michigan State University, East Lansing, MI 48824}
\affiliation{CEA-Saclay DSM/DAPNIA/SPhN, F-91191 Gif sur Yvette, France}
              \thanks{Work performed in the framework of the ESNT
              (L'Espace de Structure Nucl{\'e}aire Th{\'e}orique).}

\author{P. Bonche} 
\thanks{deceased}
\affiliation{Service de Physique Th\'eorique,
             CEA Saclay, F-91191 Gif sur Yvette Cedex,
             France}

\author{P.-H. Heenen}
\affiliation{Service de Physique Nucl\'eaire Th\'eorique,
             Universit\'e Libre de Bruxelles, Case Postale 229,
             B-1050 Bruxelles, Belgium}

\date{July 31 2006}

\pacs{21.60.-n  
      21.60.Jz, 
      21.10.-k, 
      21.10.Re  
      21.10.Ky  
      27.50.+e  
}

\begin{abstract}
We discuss results obtained in the study of shape coexistence in
the neutron-deficient \nuc{72-78}{Kr} isotopes. The method that we
used is based on the mixing of axial mean-field configurations
after their projection on particle-number and angular-momentum.
The calculations are performed with a Skyrme interaction and a
density-dependent pairing interaction. While our calculation
reproduces qualitatively and quantitatively many of the global
features of these nuclei, such as coexistence of prolate and
oblate shapes, their strong mixing at low angular momentum, and
the deformation of collective bands, the ordering of our
calculated low-lying levels is at variance with experiment. We
analyse the role of the single-particle spectrum of the underlying
mean-field for the spectrum of collective excitations.
\end{abstract}
\maketitle
%
%
\section{Introduction}
In a mean-field framework \cite{RMP}, the presence of low-lying
$0^+$ states in the spectrum of an even-even nucleus is usually
interpreted as the manifestation of shape coexistence \cite{Woo92a}:
each of the $0^+$ states, including the ground state, corresponds to
a mean-field configuration of a different shape. Nuclei around the
neutron deficient Kr and Sr have very early been considered as among
the most favorable ones for the presence of shape coexistence. First
studies were performed for $^{72}$Kr and a few neighboring odd
nuclei with the help of the  Nilsson-Strutinsky approach
\cite{Dic72a}. Detailed calculations have been carried out since
then with an improved microscopic-macroscopic model \cite{Naz85a} and
with self-consistent mean-field models using non-relativistic Skyrme
\cite{Bon85a} and Gogny \cite{Gir89a} interactions, as well as
relativistic Lagrangians \cite{Lal95a}. They confirm the presence of
oblate and prolate minima in the deformation energy surface of some
light Kr, Sr and Zr isotopes.

The existence of nearly-degenerate structures corresponding to
different deformations raises immediately the question of their
stability against dynamical effets beyond a mean-field approach. One
can expect that the physical states result in fact from a mixing of
states with spherical, prolate and oblate deformations. Such mixings
were obtained from models with parameters specifically adjusted to
the data:  the proton-neutron Interacting Boson Model (IBA-2)
\cite{Kau79a}, and a Bohr-Hamiltonian calculation built on a
microscopic-macroscopic model \cite{Pet83a}.

An alternative to a mean-field description of shape coexistence is
given by the shell model. However, the number of active particles
and holes necessary to describe the neutron-deficient Kr isotopes is
prohibitively large, and this mass region is out of reach of the
conventional shell model. The problem is tractable in models
performing a clever truncation of the configuration space as in the
complex excited VAMPIR approach using a phenomenologically modified
nuclear matter $G$ matrix as residual interaction
\cite{Pet00a,Pet02a}, or in the Shell-Model Monte Carlo (SMMC)
method. A SMMC calculation using a schematic pairing+quadrupole
interaction is presented in Ref.\ \cite{Lan03a}.

First experimental evidence for ground-state deformation in
neutron-deficient Kr isotopes was found in Ref.\ \cite{Pie81aE}.
Subsequent experiments revealed the rich structure of the low-lying
excitation spectrum in these nuclei, with coexisting and mixed bands
in \nuc{72-78}{Kr} \cite{Cha97aE,Bec99aE,Bou03aE}. Many more data on
transition probabilities, both in and out of bands, have also now
become available
\cite{Gia95aE,Ang97aE,Alg00aE,Jos02aE,Dob05aE,Goe05aE,Gad05aE,Dob05bE,Cle05aE,Bec06aE,Dha06aE}.

The description of such nuclei is a challenge for nuclear structure
models. In general, their excitation spectrum varies very rapidly
along a given isotopic line. For this reason, they constitute a
testing ground for beyond-mean-field approaches built on
Skyrme-Hartree-Fock \cite{Bon91a,Hee93a}. In this paper, we present
an application of an angular momentum projected generator coordinate
method to the description of the low-energy excitations in
\nuc{72-78}{Kr}. At present, our implementation of the method
is limited to axial mean-field states which are invariant under
spatial inversion and time reversal.
Applications of the same method to shape
coexistence in neutron-deficient Pb isotopes were presented in
Refs.\ \cite{Dug03pb,Ben03pb}. A good qualitative agreement with the
data was obtained, in particular for the relative position of the
coexisting bands which result from a mixing of oblate, spherical and
prolate configurations. However, the mixing is extremely sensitive
to many details of the model and on the effective interaction,
preventing a detailed quantitative agreement with the data. Compared
to the neutron-deficient Pb isotopes, the description of the
neutron-deficient Kr isotopes looks simpler. The coexisting bands
result from the mixing of only two kinds of structures, oblate and
prolate. The number of spherical $j$ shells which contribute to the
shell structure is smaller than for Pb, and the shells are the same
for neutrons and protons. It is therefore easier to relate the
collective states to the underlying mean-field.

Our calculated values presented below have already been used for
comparison in the experimental report of Refs.\
\cite{Goe05aE,Cle05aE}.
%
%
\section{The method}
The starting point of our method is a set of axial HF+BCS wave
functions. They are generated by self-consistent mean-field
calculations on a three-dimensional mesh in coordinate space
\cite{Bon85a,Bon05a}, with a constraint on a collective
coordinate, the axial quadrupole moment $\langle Q_{20} \rangle$.
In a spherical nuclear shell model picture, such mean-field states
incorporate particle-particle correlations through pairing and
many-particle many-hole correlations through nuclear deformation.
As a result, the mean-field states break several symmetries of the
exact many-body states. These symmetry violations make difficult
the connection between mean-field results, expressed in the
intrinsic frame of reference of the nucleus, and spectroscopic
data, obtained in the laboratory frame of reference. This
motivates the second step of our method, the restoration of the
symmetries associated with particle numbers and rotation. Another
ambiguity in the interpretation of mean-field results arises when
the deformation energy varies slowly as a function of a shape
degree of freedom, in particular when a potential energy surface
presents several minima separated by a low barrier. In such a
case, to assign a physical state to each minimum is not a
well-founded approximation. This problem is eliminated by the third
step of our method: for each angular momentum, states obtained by
projecting mean-field configurations of different deformation are
used as the generating functions of the generator coordinate
method (GCM). The weight coefficients of the mixing are determined
by varying the energy and by solving the Hill-Wheeler-Griffin
equation \cite{Hil53a}. The configuration mixing
removes the contribution of vibrational excitations from the
ground state wave function. It permits also to construct a
spectrum of excited states. A detailed introduction to the method
has been given in Ref~\cite{Ben05b}.

The same effective interaction is used to generate the mean-field
states and to perform the configuration mixing: the Skyrme
interaction SLy6 \cite{SLyx} in the particle-hole channel and a
density-dependent, zero-range interaction in the pairing channel. As
required by the SLy6 parametrization, the full two-body
center-of-mass correction is included in the variational equations
to generate the mean-field and in the calculation of the projected
GCM energies. The strength of the pairing interaction is the same as
in previous studies of light and medium-mass nuclei, i.e.\ $-1000$
MeV fm$^3$ for protons and neutrons in connection with cutoffs above
and below the Fermi energy, as defined in Ref.~\cite{Rig99}. Similar
implementations of angular-momentum projected GCM have been
developed also for the non-relativistic Gogny interaction
\cite{Egi04a} and for relativistic mean-field Lagrangians
\cite{Nik06a}.

To define a dimensionless deformation parameter, we use the sharp
edge liquid drop relation between the axial quadrupole moment
$Q_2^{(i)}$ and a parameter $\beta_2^{(i)}$
\begin{equation}
\label{eq:beta}
\beta_2^{(i)}
= \sqrt{\frac{5}{16 \pi}} \, \frac{4 \pi Q_2^{(i)}}{3 R^2 A}
,
\end{equation}
with \mbox{$R = 1.2 \, A^{1/3}$ fm}. The mass dependence of the
quadrupole moment is partly removed in this parameter $\beta_2$. It
should not be compared with the multipole expansion parameters that
are used in microscopic-macroscopic models whose origin is not the
same and usually are significantly smaller.

With our method, $B(E2, J \to J')$ values for in-band and
out-of-band transitions as well as spectroscopic multipole moments
$Q_s (J)$ are determined directly in the laboratory frame of
reference~\cite{Ben03a}. As we use the full model space of
occupied states, we neither have to distinguish between valence
particles and a core, nor is there a need for effective charges.

To make a comparison with other approaches easier, it is useful to
define quantities similar to intrinsic frame deformations from the
spectroscopic and transition moments. An intrinsic transition charge
quadrupole moment $Q^{(t)}$ can be derived from the $B(E2)$ values
within the static rotor model
\begin{equation}
\label{eq:Qt}
Q^{(t)}(J)
= \sqrt{\frac{16 \pi}{5}
  \frac{B(E2,J \to J-2)}
       {( J \, 0 \, \, 2 \, 0 |  J-2 \, 0 )^2 e^2}}
.
\end{equation}
In the same way, an intrinsic charge quadrupole moment can be
related to the spectroscopic quadrupole moment $Q_c(J)$ via the
relation
\begin{equation}
\label{eq:Qs}
Q^{(s)}(J)
= \frac{(J+1)(2J+3)}{3 K^2 - J(J+1)} \, Q_c (J)
.
\end{equation}
For axially-symmetric shapes, i.e.\ pure \mbox{$K=0$} states, this
relation reduces to $Q^{(s)}(J) = - Q_c(J) \; (2J+3)/J $, with a
change of sign between $Q^{(s)}$ and $Q_c(J)$. Dimensionless
deformation parameters $\beta_2^{(t)}$ and $\beta_2^{(s)}$ can be
determined from $Q^{(t)}$ and $Q^{(s)}$, respectively, through an
equation similar to Eq.\ (\ref{eq:beta}), with $A$ replaced by $Z$.
Within a given band, $Q^{(s)}$ and $Q^{(t)}$ might still depend
strongly on angular momentum.

$Q^{(s)}$ and $Q^{(t)}$ measure different properties: $Q^{(s)}$
depends on a single state, while $Q^{(t)}$ probes the geometry of
the initial and final states whose wave functions can correspond to
very different mixings of projected mean-field states. In general,
these two moments take different values. Their near equality
indicates that the assumptions of the static-rotor model are
fulfilled, i.e.\ a well-deformed rotational band not mixed with
other bands.

%
%
\section{Results}
%
%
\subsection{Potential landscapes}
\begin{figure}[t!]
\includegraphics{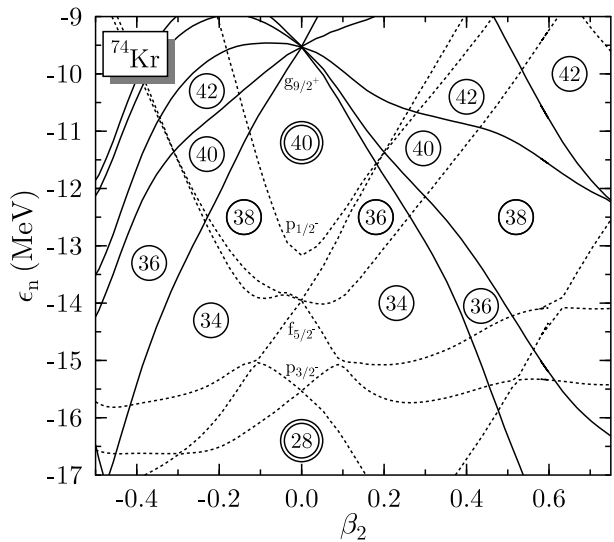}
\caption{\label{fig:kr74:spe}
Nilsson plot of neutron single-particle energies with positive (solid lines)
and negative (dotted lines) parity as a function of the intrinsic mass
deformation $\beta_2^{(i)}$, as obtained for \nuc{74}{Kr}.
}
\end{figure}

Figure \ref{fig:kr74:spe} provides the Nilsson diagram of the
self-consistent single-particle energies for neutrons. The diagram
for protons looks very similar, but shifted in energy. One can see
two main differences with the single-particle energies used in some
other methods. In the Woods-Saxon and the Folded-Yukawa potentials
used in the microscopic-macroscopic calculations of
Ref.~\cite{Naz85a} and \cite{Mol97a} respectively, as in the
Woods-Saxon potential used to generate the single-particle spectrum
for the SMMC calculations of Ref.~\cite{Lan03a}, the $p_{3/2}$ and
$p_{1/2}$ levels are closer to the $g_{9/2}$ level by approximately
2 MeV and the separation between the $f_{5/2}$ and the $g_{9/2}$
levels is larger. As a result, the $p_{3/2}$ and the $f_{5/2}$
levels are much closer and their order is even changed in the case
of the latter two models. These modifications have a strong effect
on the deformed gaps which may correspond to quite different
deformations and vary in size. The spin-orbit splitting for the $f$
and $p$ levels in the three potentials are very similar to ours. The
differences between the single particle spectra must thus be related
to the relative position of states with different orbital angular
momentum within a given shell.

\begin{figure}[t!]
\includegraphics{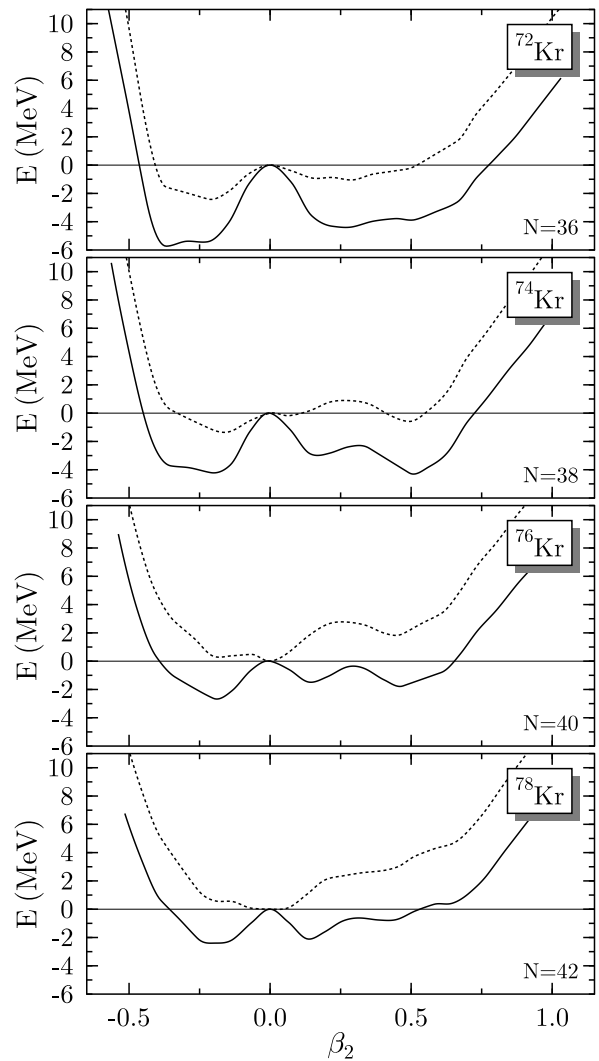}
\caption{\label{fig:all:pes}
Mean-field (dotted) and \mbox{$J=0$} projected deformation energy
curves (solid) for \nuc{72-78}{Kr} (see text).
}
\end{figure}

Figure \ref{fig:all:pes} shows the mean-field and \mbox{$J=0$}
projected energy curves for \nuc{72-78}{Kr} obtained with the SLy6
Skyrme parametrization. Throughout this paper, the projected
energy is plotted as a function of the intrinsic $\beta_2^{(i)}$
value of the mean-field states from which they are obtained. In
our opinion, this quantity provides the most convenient and
intuitive label that can be defined for all states, irrespective
on the level of modelling. However, it should not be
misinterpreted as an observable. With our method, one calculates
transition and spectroscopic multipole moments in the laboratory
frame, which can be directly compared to experimental data.
However, the spectroscopic moments which characterize a state do
not provide useful coordinates to plot potential energy curves as
they scale with angular momentum and are even identically zero for
$J=0$. For projected states with \mbox{$J>0$} and large intrinsic
deformation in nuclei with A larger than 100, $\beta_2^{(i)}$ is
very close to the intrinsic deformation $\beta_2^{(s)}$ determined
from the laboratory frame quadrupole moment $Q^{(s)}$ through the
static rotor model, Eq.\ (\ref{eq:Qs}). Note that any coordinate
might be misleading for the GCM, as the metric is related to the
inverse of the overlap matrix, which has no direct connection to
any deformation.

The mean-field energy landscapes (dotted lines in Fig.\
\ref{fig:all:pes}) show that the energies of the four nuclei vary
very slowly with deformation. Our calculations predict also a
transition from a nucleus with coexisting prolate and oblate minima
in \nuc{72}{Kr} to a soft, spherical, anharmonic vibrator in
\nuc{78}{Kr}. The many shallow local minima and plateaus in the
total energy curves can be directly related to the gaps in the
Nilsson diagram in Fig.\ \ref{fig:kr74:spe}.

The two minima in the mean-field energy curve (dotted line) of
\nuc{74}{Kr} reflect the \mbox{$N=38$} gaps in the Nilsson diagram
at small oblate and large prolate deformations. For \nuc{76}{Kr},
the spherical mean-field minimum is related to the large spherical
\mbox{$N=40$} subshell closure, while the shallow oblate and
prolate structures correspond to two deformed \mbox{$N=40$} gaps
in the Nilsson diagram. The prolate minimum has moved to smaller
deformation than in \nuc{74}{Kr}, the deformed \mbox{$N=38$} gap
being at larger deformations than the \mbox{$N=40$} gap. The
spherical \mbox{$N=40$} subshell closure is strong enough to
stabilize the spherical shape up to the \mbox{$N=42$} isotope
\nuc{78}{Kr}.

As can be seen from the solid lines in Fig.\ \ref{fig:all:pes}, the
energy landscapes are qualitatively modified by the projection on
\mbox{$J=0$}. Since a spherical mean-field state is already a
\mbox{$J=0$} state, the energy gain by projecting it on angular
momentum is zero while the projection of a deformed
mean-field state always leads to an energy gain, which increases
very rapidly at small deformation. In almost all spherical and soft
nuclei, this creates minima at prolate and oblate deformations with
$|\beta_2^{(i)}|$ values around 0.1. These states usually have a
large overlap close to 1, which means they are not two different
states, but represent the same one, which can be associated with a
``correlated spherical state''. In nuclei with shallow mean-field
minima at small deformations as the Kr isotopes, the projection
merges this near-spherical spherical minimum with the slightly
oblate one into a broad structure. The prolate mean-field minimum
being at larger deformation, two distinct structures appear on the
prolate side in \nuc{74-78}{Kr}, a ``spherical'' one at a $\beta_2$
value around 0.15, and a well-deformed one at a $\beta_2$ value
around 0.5.

The topology of the \nuc{72}{Kr} energy surface on the oblate side
can hardly be explained by a single large oblate \mbox{$N=36$} gap
in the Nilsson diagram. The mean-field and $J=0$ minima at large
oblate deformations seem to result from a delicate balance of shell
effects from $N=34$ to $N=40$ which reduce the level density around
the Fermi surface sufficiently to create a shallow oblate minimum
and a very soft energy surface. The shallow minimum on the prolate
side can be associated with two close \mbox{$N=36$} gaps.

The fact that the structures in the potential landscapes can be
easily associated with gaps in a Nilsson diagram  is an attractive
feature of our approach which makes straightforward the connection
with simpler models. This is an advantage compared to the
interacting shell model, in which states are constructed as
non-intuitive $n$p-$n$h states in a spherical basis.

One can wonder whether it is meaningful to restrict the description
of the light Kr isotopes to axial shapes. There have been a few
explorations of the triaxial degree of freedom which can help to
answer this question. Yamagami \etal\ \cite{Yam01a} have calculated
the potential landscape of \nuc{72}{Kr} along several shape degrees
of freedom, including non-axial octupole deformations, using the
Skyrme SIII interaction. Their potential energy curve is similar to
ours for purely quadrupole axial deformations. They found that the
prolate and oblate energy minima are separated by a triaxial barrier
of 500 keV, and that the oblate shapes are soft with respect to
non-axial octupole deformations. Bonche \etal\ \cite{Bon85a} have
obtained rather large triaxial barriers for \nuc{74}{Kr} and
\nuc{76}{Kr}, also using the SIII interaction. Let us also mention
that Almehed and Walet \cite{Alm04a} have self-consistently
determined a collective path in \nuc{72}{Kr} for different values of
the angular momentum in a small model space using a schematic
interaction. For \mbox{$J=0$}, they obtain a purely axial path that
connects the oblate ground state and the prolate minimum. For
\mbox{$J=2$}, the path that they obtain does not cross the spherical
configuration and makes a small excursion through triaxiality, as
can be expected from the projected energy curves. From these studies
of triaxiality, one can conclude that in most models, the oblate and
prolate minima found as a function of the axial quadrupole moment
are true minima. The barrier between the minima in the $\gamma$
degree of freedom may not be very high, but to restrict the model to
axial symmetry can be expected to provide a valid first approximation.

%
%
\subsection{Spectroscopy}

We first focus on the three isotopes for which very complete sets
of experimental data are available. This enables us to introduce
our notations and all the ingredients which are necessary for a
comparison between our results and the data. We will come back to
the more prospective case of \nuc{72}{Kr} at the end.

%
\subsubsection{\nuc{74}{Kr}}
\begin{figure}[t!]
\includegraphics{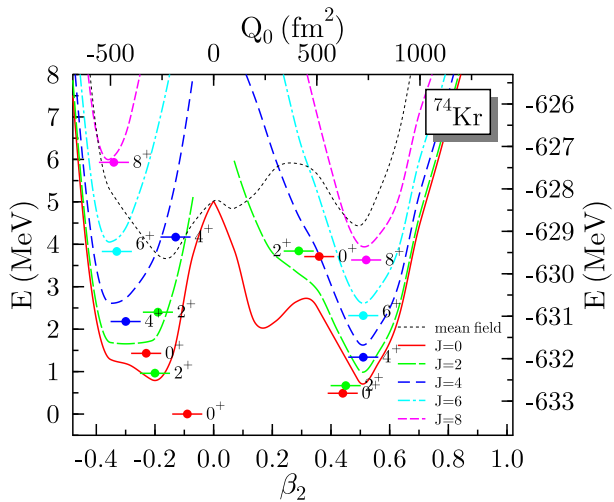}
\caption{\label{fig:kr74:ex} (color online). Mean-field energy
curve, angular-momentum projected energy curves and projected GCM
for \nuc{74}{Kr}. The projected energy curves are plotted as a
function of the intrinsic deformation of the mean-field they are
projected from, the projected GCM levels at the average deformation
$\bar\beta_2$, Eq.\ (\ref{eq:bar:beta}). The labels on the
right-hand side give the total binding energy as calculated, the
labels on the left-hand side the energy relative to the \mbox{$J=0$}
GCM ground state. }
\end{figure}

\begin{figure}[t!]
\includegraphics{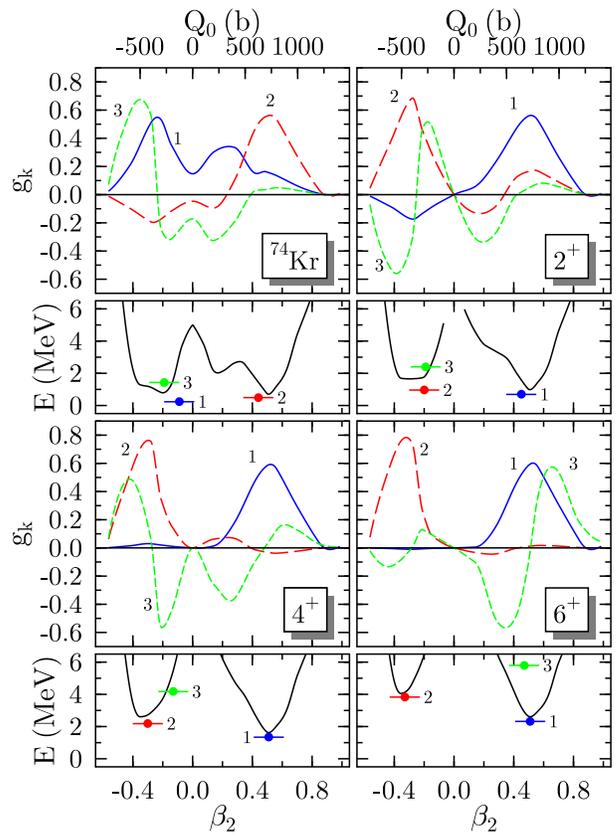}
\caption{\label{fig:kr74:wf}
(color online). Collective wave functions $g_k^{(J)}$ for the
lowest states with \mbox{$J=0$}, 2, 4, and 6 for \nuc{74}{Kr} as a
function of the intrinsic deformation of the mean-field states
they are constructed from. The corresponding energies and average
deformations $\bar\beta_2$ are plotted together with the projected
energy surface. }
\end{figure}

In Fig.\ \ref{fig:kr74:ex} are given the lowest projected GCM levels
for \mbox{$J=0$} up to 8 together with the corresponding projected
energy curves. The collective levels are plotted at the mean
deformation $\bar\beta_2^{(i)}$ of the mean-field states from 
which they are constructed, defined as
\begin{equation}
\label{eq:bar:beta}
\bar\beta_2^{(i)}
= \sum_{\beta_2^{(i)}} \; \beta_2^{(i)} \; g_{J,k}^2(\beta_2^{(i)})
,
\end{equation}
which provides in many cases an intuitive picture of the band
structure in a nucleus. Figure \ref{fig:kr74:wf}, however,
illustrates the limits of the meaning of $\bar\beta_2$. In this
figure are shown the collective wave functions $g_i^J$ for
\mbox{$J=0$}, 2 and 4. All low-lying $0^+$ states result from a
strong mixing between prolate and oblate mean-field configurations.
The values of $\bar\beta_2$ for the ground state and the second
excited state reflect the dominance of oblate and prolate
deformations. For the ground state $0^+$ state, the very small value
of $\bar\beta_2$ does not mean that this state is nearly spherical,
but that the weights of prolate and oblate shapes are nearly equal.
For higher $J$ values, the mixing between oblate and prolate
configurations is less pronounced and the value of $\bar\beta_2$
represents better the structure of the states.

As could be seen in Fig.\ \ref{fig:all:pes}, the \mbox{$J=0$}
projected energy curve for \nuc{74}{Kr} presents two nearly
degenerate minima at oblate and large prolate deformations, and a
third shallow excited minimum at small deformation. This structure
is reflected in the spectrum of GCM states plotted in Fig.\
\ref{fig:kr74:ex} and in the wave functions drawn in Fig.\
\ref{fig:kr74:wf}.

As can be seen in Fig.~\ref{fig:kr74:wf}, the \mbox{$J=0$} GCM wave
functions strongly mix projected oblate, near-spherical and prolate
configurations. The near degeneracy of the oblate and prolate minima
is lifted, the first excited $0^+$ state having an excitation energy
of 0.49 MeV. The situation becomes simpler for larger angular
momenta. The first $2^+$ level still strongly mixes prolate and
oblate deformations and has a wave function rather similar to the
first excited $0^+$ state. On the contrary, the second $2^+$ state
is mainly constructed from oblate mean-field configurations
projected on \mbox{$J=2$} with a negative mean deformation larger in
modulus than that of the ground state. The two first $4^+$ and $6^+$
states are dominated by only prolate or oblate deformations. One can
thus expect that an approximation based on a fixed mean-field
configuration, like the cranked mean-field methods, is justified
beyond \mbox{$J=4$}.

The nodal structure of the collective wave function is more
complicated than the structure that one would obtain in a
one-dimensional potential well. It reflects the fact that the
rotation of a deformed wave function around the $y$-axis generates
wave functions which cannot be represented along a single axis: the
exchange of the $x$ and $z$ axes generates prolate and oblate shapes
for two different values of the angle $\gamma$. For this reason, the
angular-momentum projected GCM wave functions is not
one-dimensional, even when only axial states are mixed
\cite{Ben03a}.

\nuc{74}{Kr} provides an excellent example that a comparison between
the intrinsic deformation of a mean-field minimum and the transition
quadrupole moment derived from a $B(E2)$ value might be misleading.
There are three reasons for that: (i) for light and medium-mass
nuclei, the minima of the projected energy curve are usually at
larger prolate and oblate deformations than the minima of the
mean-field energy curve, (ii) the minima of the projected
\mbox{$J=0$} and \mbox{$J=2$} energy curves correspond to the
projection of mean-field states with different deformation and the
assumption of a static rotor is not valid at low angular momentum,
and (iii) the wave functions of the low-lying states are quite broad
and strongly mix oblate or prolate configurations; moreover, the
mixing is different for the $0^+$ states and $2^+$ states.

A schematic two-level-mixing model has been used by Becker \etal\
\cite{Bec99aE,Kor01aP} to analyze the experimental data. In this
model, the energies of "pure" prolate and oblate configurations are
deduced from an extrapolation to low spins of the high spin parts of
the bands. This is valid if configuration mixing decreases with
spin, which seems reasonable in view of our results. Data are
described well by assuming that the unperturbed levels are nearly
degenerate with only a 20 keV energy difference. This leads to
nearly equal contributions from the oblate and prolate unperturbed
states. This result is consistent with our calculations, although
the assumption that one has to mix only two configurations is too
crude. The complicated structures that we obtain for the collective
wave functions, of Fig.\ \ref{fig:kr74:wf} cannot be reduced to the
mixing of two or even three configurations with a well defined
shape. It does not seem therefore possible to connect our results
with a phenomenological two-level mixing model.

\begin{figure}[t!]
\includegraphics{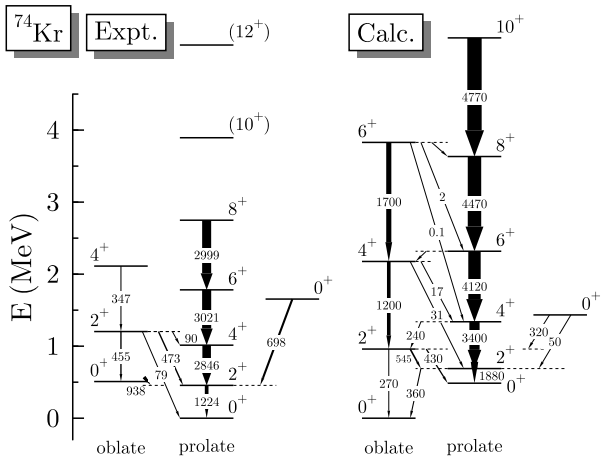}
\caption{\label{fig:kr74:collspec}
Spectrum of collective states for \nuc{74}{Kr} as seen in
experiment (left) and as obtained from our calculation (right).
The arrows with numbers denote the $B(E2, J \to J-2)$ values for
transitions between states with different $J$ and $B(E2, J \to J)$
for transitions between states with the same $J$, both given in
$e^2$ fm$^{4}$. Experimental values for the $B(E2)$ values are
taken from \cite{Goe05aE,Cle05aE}. The labels "oblate" and
"prolate" given to the bands correspond to the main components of
the collective wave functions. }
\end{figure}

Theoretical and experimental excitation spectra and $B(E2)$ values
are compared in Fig.\ \ref{fig:kr74:collspec}. As in the experiment,
the theoretical levels are ordered in bands on the basis of their
spectroscopic quadrupole moments and of the dominant $E2$
transitions. A low-lying $\gamma$ band, which has a $2^+$ band head
at 1.74 MeV, has been omitted since \mbox{$K=2$} states are outside
our model space. The $B(E2)$ values are always given for transitions
from $J \to J-2$ or from $J \to J$, although values found in the
literature correspond sometimes to $B(E2,J \to J+2)$ values, which are
by definition larger by a factor 5. Our calculation is only partly 
consistent with
experiment. We reproduce the coexistence of oblate and prolate
bands, strongly mixed at low angular momentum, but the lowest $0^+$
states do not have the right ordering. Our calculated ground state
is predominantly oblate, but the prolate band becomes yrast already
at \mbox{$J = 2$}. For higher spins the spectrum is too stretched
out, a problem that has already been encountered in previous GCM
studies \cite{Ben05b,Egi04a}.

The in-band transition probabilities are slightly overestimated for
both bands, which hints at a slightly too large deformation of the
dominating configurations in both bands. Nevertheless, all
calculated values are within a factor two of the experimental ones.
Out-of-band transitions between states at the bottom of
the bands are large, in particular the experimental value of  $B(E2,
2^+_{\text{pro}} \to 0^+_{\text{obl}})$ of $938^{+110}_{-91}$ $e^2$
fm$^4$ given in \cite{Cle05aE}, or $1120 \pm 460$ $e^2$ fm$^4$ given
in \cite{Bou03aE}. Our predicted out-of-band $B(E2)$ values become
weak already at \mbox{$J=4$}.

\begin{figure}[t!]
\includegraphics{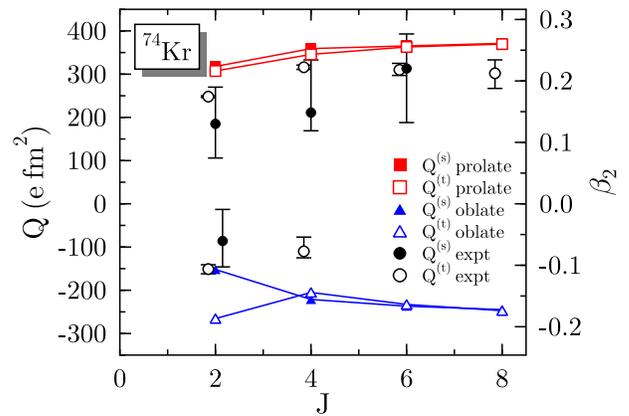}
\caption{\label{fig:kr74:Qs} (color online). Comparison of
calculated and experimental $Q^{(s)}(J)$ and $Q^{(t)}(J \to J-2)$
moments for states in the prolate and oblate bands, assuming that
all states are purely axial. Note that $-Q^{(t)}$ is plotted for
oblate states for better comparison with $Q^{(s)}$. Experimental
values are taken from \cite{Cle05aE}. }
\end{figure}

The theoretical spectroscopic quadrupole moments are compared with
experiment in Fig.\ \ref{fig:kr74:Qs}. These moments provide by
their sign the only unambiguous way to assign a prolate or an oblate
character to a band. To compare them with in-band $B(E2)$ values as
well, $Q^{(s)}$ and $Q^{(t)}$ values derived from the static rotor
model assuming \mbox{$K=0$} are shown. Except for the oblate $2^+$
state, our calculated $Q^{(s)}$ and $Q^{(t)}$ values are very
similar. Experimental values for $Q^{(s)}$ and $Q^{(t)}$ are also
similar within the large error bars of $Q^{(s)}$. Our calculation
gives systematically larger deformation for both the prolate and
oblate states as already discussed above.

%
%
\subsubsection{\nuc{76}{Kr}}

\begin{figure}[t!]
\includegraphics{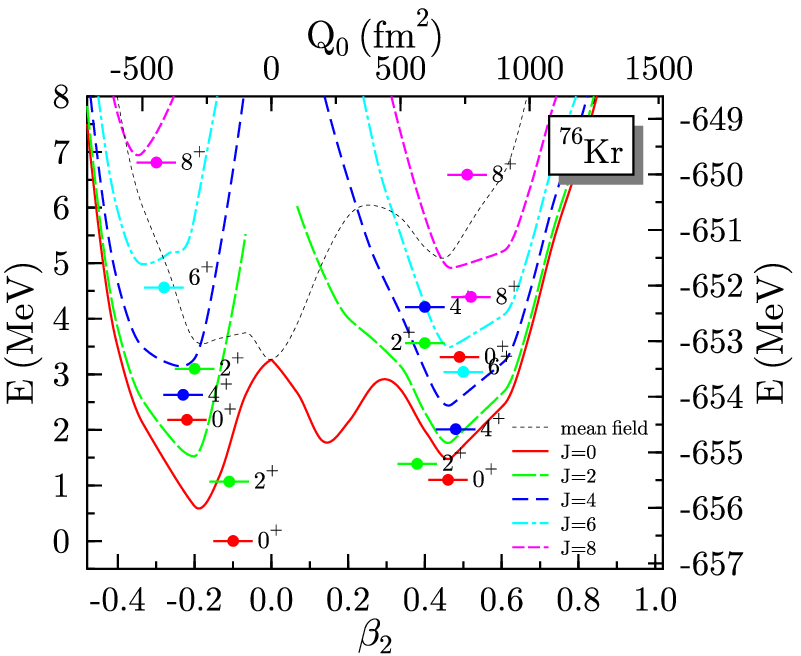}
\caption{\label{fig:kr76:ex}
(color online).
The same as Fig.\ \ref{fig:kr74:ex} for \nuc{76}{Kr}.
}
\end{figure}

\begin{figure}[t!]
\includegraphics{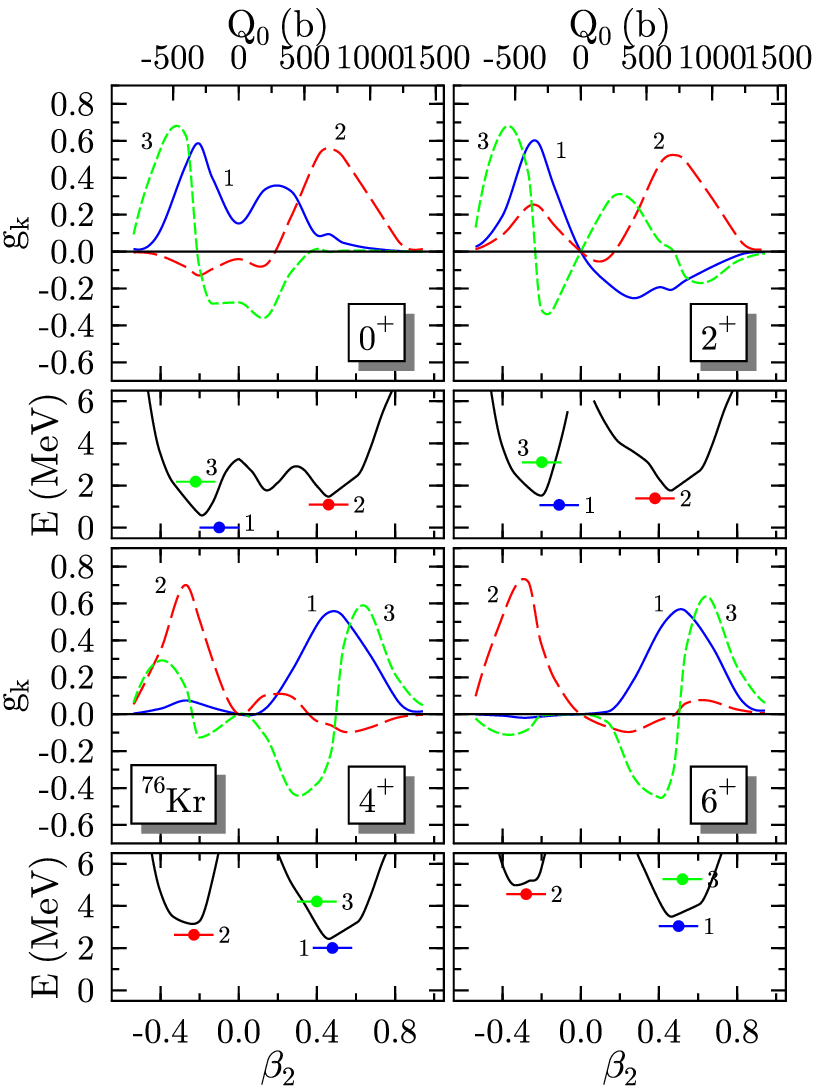}
\caption{\label{fig:kr76:wf}
(color online).
The same as Fig.\ \ref{fig:kr74:wf} for \nuc{76}{Kr}.
}
\end{figure}

\begin{figure}[t!]
\includegraphics{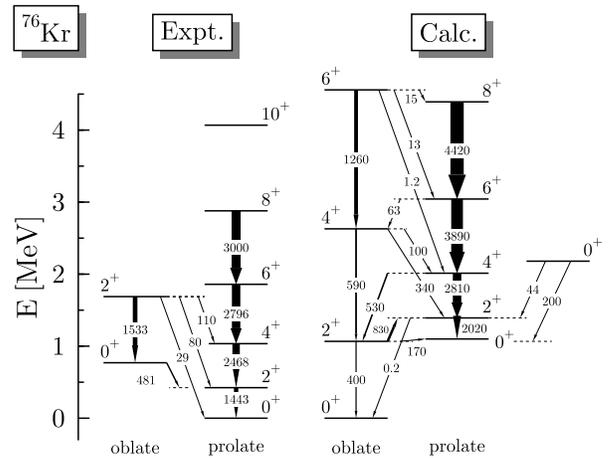}
\caption{\label{fig:kr76:collspec}
The same as Fig.\ \ref{fig:kr74:collspec} for \nuc{76}{Kr}.
Experimental data are taken from \cite{Cle05aE}.
}
\end{figure}

The projected energy curves in \nuc{76}{Kr}, plotted in
Fig.~\ref{fig:kr76:ex}, are very similar to those obtained for
\nuc{74}{Kr}, with coexisting prolate and oblate bands, except that
the prolate minima at $\beta_2$ values around 0.45 are less
pronounced. As a consequence, the ground state wave function is more
dominated by oblate configurations than that of \nuc{74}{Kr}, as
confirmed by the shape of the collective wave functions given in
Fig.\ \ref{fig:kr76:ex}. This is at variance with experiment, which
indicates a prolate ground state, see also Fig.\
\ref{fig:kr76:wf}. Again, we omit from Fig.\
\ref{fig:kr76:collspec} the experimental states assigned to a
$\gamma$ band with a $2^+$ band head at 1.222 MeV; see, e.g.\
\cite{Dob05aE,Cle05aE}. The calculated prolate band head has a
higher excitation energy than for \nuc{74}{Kr}. The prolate band
becomes yrast at \mbox{$J=4$}. Compared to \nuc{74}{Kr}, the
increased "purity" of states within the prolate and oblate bands
happens at higher angular momenta, which is reflected in the large
out-of-band $B(E2)$ values up to \mbox{$J=4$} in Fig.\
\ref{fig:kr76:collspec}.

\begin{figure}[b!]
\includegraphics{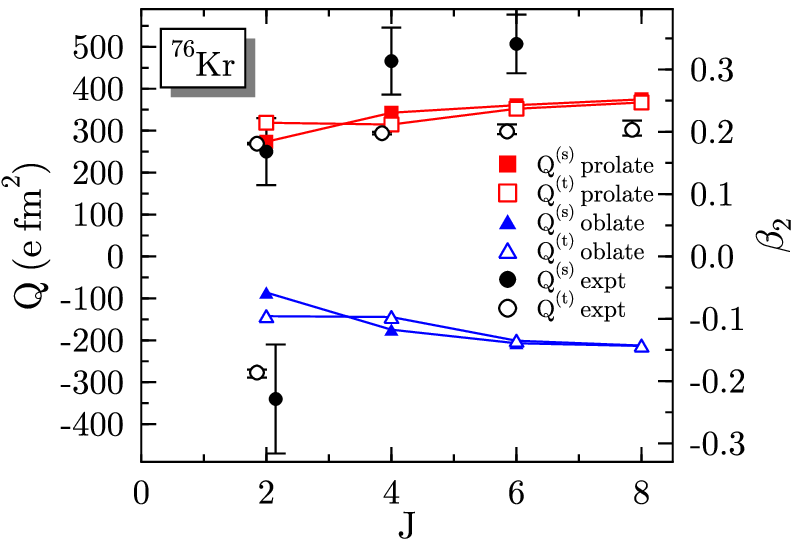}
\caption{\label{fig:kr76:Qs}
(color online).
Same as Fig.\ \ref{fig:kr74:Qs}, but for \nuc{76}{Kr}.
Experimental values are taken from \cite{Cle05aE}. }
\end{figure}

Figure \ref{fig:kr76:Qs} compares the intrinsic quadrupole moments.
As for \nuc{74}{Kr}, our calculated values for $Q^{(s)}$ and
$Q^{(t)}$ are very similar, with small deviations at low $J$.
Notably, this is not the case for the experimental values when
assuming \mbox{$K=0$}, which might hint at a small admixture of
triaxial shapes to the prolate band, which is outside the scope of
our calculation. Our calculated $Q^{(t)}$ are very close to the
experimental values for the prolate band, while the $Q^{(t)}$ of the
$2^+$ state in the oblate band is underestimated.

%
%
\subsubsection{\nuc{78}{Kr}}

\begin{figure}[t!]
\includegraphics{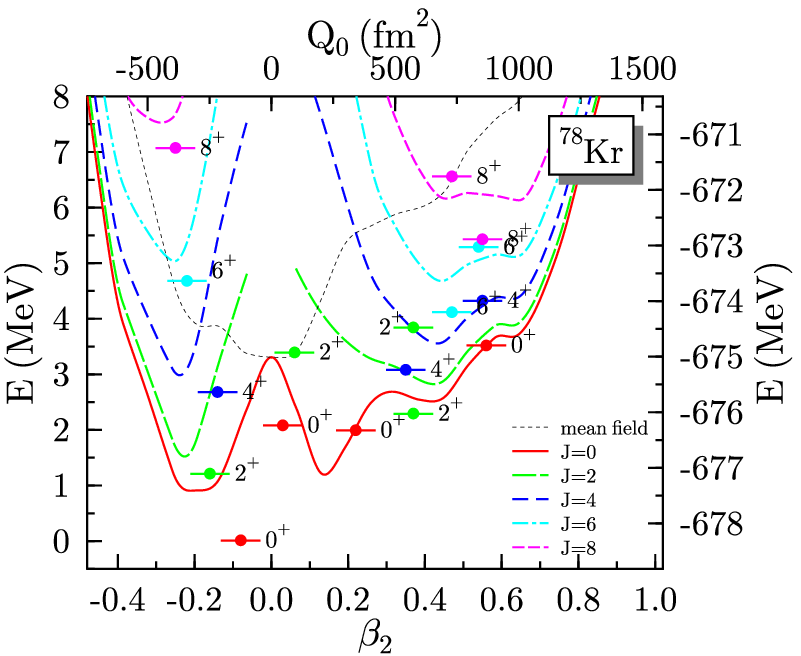}
\caption{\label{fig:kr78:ex}
(color online).
Same as Fig.\ \ref{fig:kr74:ex}, but for \nuc{78}{Kr}. }
\end{figure}

\begin{figure}[t!]
\includegraphics{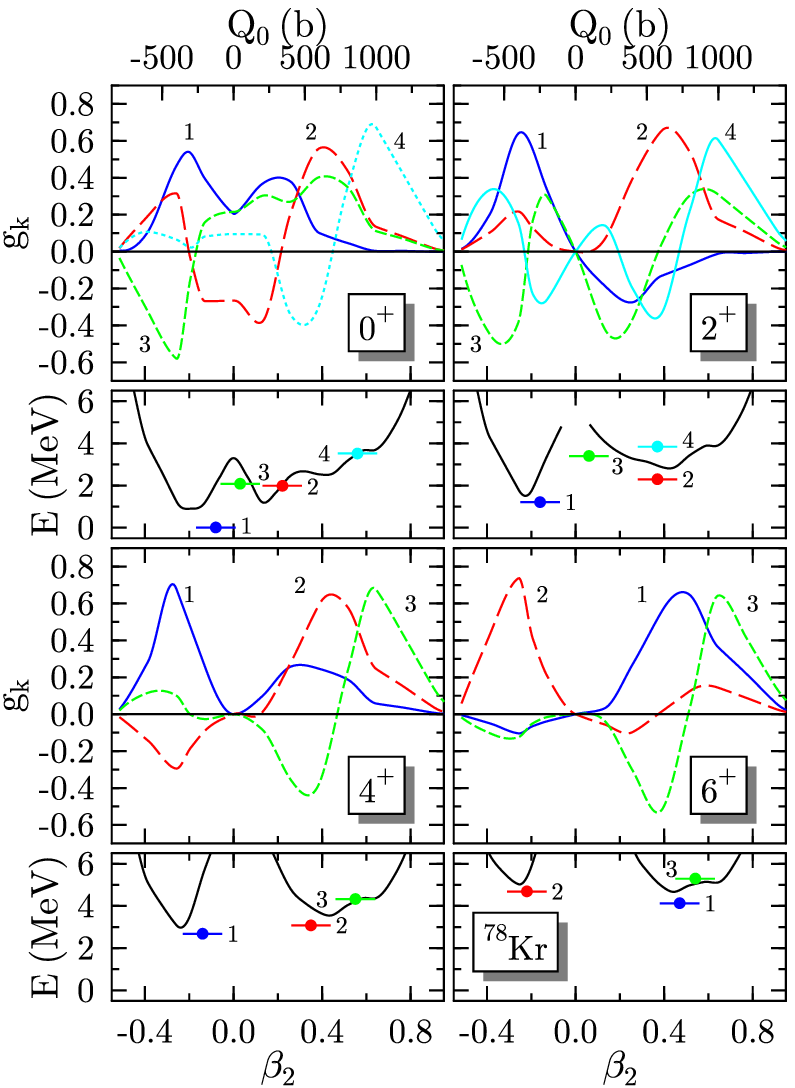}
\caption{\label{fig:kr78:wf}
(color online).
Same as Fig.\ \ref{fig:kr74:wf}, but for \nuc{78}{Kr}. }
\end{figure}

Among the Kr isotopes, \nuc{78}{Kr} has the lowest excitation energy
for the first $2^+$ state, and the largest $B(E2, 2^+_1 \to 0^+_1)$
value, corresponding to a transition quadrupole deformation
$\beta_2^{(t)}$ of 0.35. The rotational band built on the ground
state has been seen up to \mbox{$J=24$}. Two additional bands are
known at low excitation energy, a first usually interpreted as a
$\gamma$ band built on a low-lying $2^+$ state at 1.148 MeV, and a
second one built on a $0^+$ state at 1.017 MeV, originally assumed
to be oblate. According to the recent analysis of \cite{Bec06aE},
the spectroscopic quadrupole moment of the $2^+$ state in this band
is negative, which indicates that this band should, in fact, be
prolate.

\begin{figure}[t!]
\includegraphics{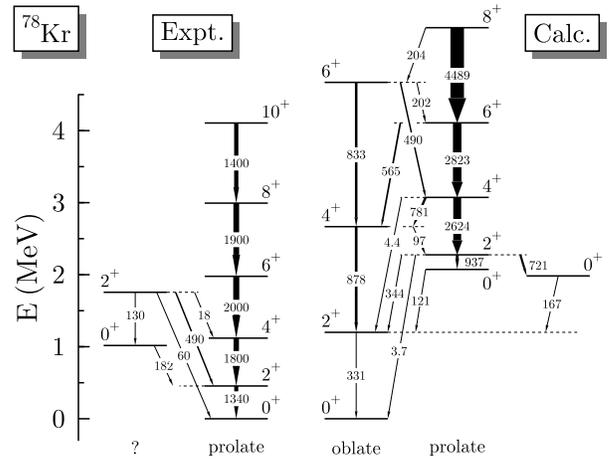}
\caption{\label{fig:kr78:collspec}
The same as Fig.\ \ref{fig:kr74:collspec}, but for \nuc{78}{Kr}.
Experimental values for the $B(E2)$ values are taken from
\cite{Bec06aE}.
}
\end{figure}

Looking to the energy curves of \nuc{78}{Kr}, the collective
states obtained from the projected GCM, Fig.\ \ref{fig:kr78:ex},
and the collective wave functions,  Fig.\ \ref{fig:kr78:wf}, one
can attribute to the ground state a dominating oblate structure,
and to the band head of the first excited band, located at 2 MeV,
a well-deformed prolate shape.  All low-lying states are strongly
mixed, which leads to large out-of-band $B(E2)$ values up to
\mbox{$J=6$}.

Data for transition moments are taken from a recent Coulomb
excitation experiment \cite{Bec06aE}. Values for in-band
transitions in the yrast band were published earlier in Refs.\
\cite{Jos02aE,Dha06aE}, with a $B(E2; 2^+_1 \to 0^+_2)$ value of
91(5) e$^2$ fm$^4$ in \cite{Gia95aE}, compatible with the value of
130(40) obtained in \cite{Bec06aE}.

\begin{figure}[t!]
\includegraphics{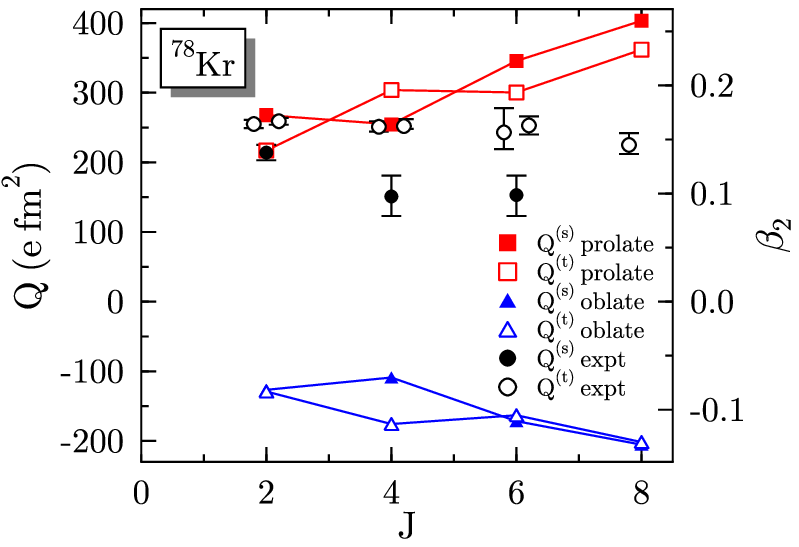}
\caption{\label{fig:kr78:Qs}
(color online).
Same as Fig.\ \ref{fig:kr74:Qs}, but for \nuc{78}{Kr}.
Experimental values are taken from \cite{Bec06aE,Jos02aE}. }
\end{figure}

As can be seen in Figs.\ \ref{fig:kr78:collspec} and
\ref{fig:kr78:Qs}, our values are close to the experimental data for
the low-lying states in the prolate band. The slight increase of
$Q^{(t)}$ with $J$ that we obtain can be related to the gradual
disappearance of the prolate minimum at $\beta_2$ values around 0.45
and to the deepening of the shoulder at $\beta_2$ values around 0.6,
which  becomes the prolate minimum at \mbox{$J \geq 8$}, see Fig.\
\ref{fig:kr78:ex}. In contrast, the experimental $Q^{(t)}$'s are
constant or slightly decreasing with $J$, see also
\cite{Jos02aE,Dha06aE}.

The large  mixing between the prolate $0^+$ state with another
excited $0^+$ state considerably reduces the in-band $B(E2;
2^+_{\text{pro}} \to 0^+_{\text{pro}})$ value, as the $E2$ strength
from the $2^+_{\text{pro}}$ level is nearly equally distributed.
Therefore, our $B(E2; 2^+_{\text{pro}} \to 0^+_{\text{pro}})$
probability is slightly smaller than the experimental $B(E2; 2^+_1
\to 0^+_1)$ from the corresponding experimental state. As a
consequence of this strong mixing, the calculated $Q^{(s)}$ and
$Q^{(t)}$ values differ more than for \nuc{74}{Kr} and \nuc{76}{Kr},
$Q^{(s)}$ being larger than $Q^{(t)}$ except for \mbox{$J=4$}.
Experimentally, $Q^{(s)}$ values are  smaller than $Q^{(t)}$ ones
\cite{Bec06aE}.

%
%
\subsubsection{\nuc{72}{Kr}}

Rotational bands at high spin in \nuc{72}{Kr} have been
investigated extensively, mainly motivated by the quest for
fingerprints of \mbox{$T=0$} proton-neutron
pairing~\cite{FLB01a,FLB03a}. Much less is known about the
structure of this nucleus at low excitation energy. Only one band
and two states that cannot be grouped into bands have been
observed at low spin \cite{Bou03aE}. The precise structure of the
low-spin yrast states is not completely clear either, as the
ground state is argued to be oblate, in parts by consistency with
theoretical predictions, while at higher spin the yrast states are
assumed to be prolate. It was also argued in Ref.\ \cite{Ang97aE}
that the large $Q^{(t)}$ value obtained from the $B(E2; 8^+ \to
6^+) = 2090 (780)$ $e^2$ fm$^4$ favors its interpretation as a
transition between prolate states. \footnote{The authors of Ref.\
\cite{Ang97aE} use a definition of $\beta_2$ different from ours,
so that the values cannot and should not be compared. } We obtain,
however, a much larger $B(E2)$ for the $8^+_{\text{pro}} \to
6^+_{\text{pro}}$ transition than the experimental value, which is
in fact consistent with our $8^+_{\text{obl}} \to
6^+_{\text{obl}}$ value. In view of the systematic overestimation
of the experimental $B(E2)$s at large spins given by our method
for the heavier Kr isotopes, one cannot draw a conclusion on the
nature of the $8^+$ state in \nuc{72}{Kr}. However, it is clear
that these large $B(E2)$ values do not exclude an oblate band as
previously argued.

\begin{figure}[t!]
\includegraphics{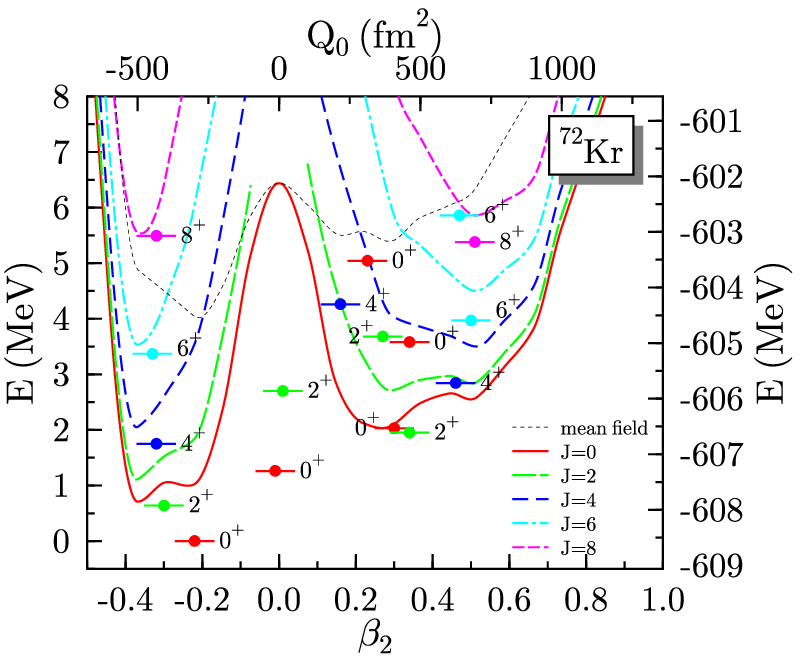}
\caption{\label{fig:kr72:ex}
(color online).
Same as Fig.\ \ref{fig:kr74:ex} for \nuc{72}{Kr}. }
\end{figure}

\begin{figure}[t!]
\includegraphics{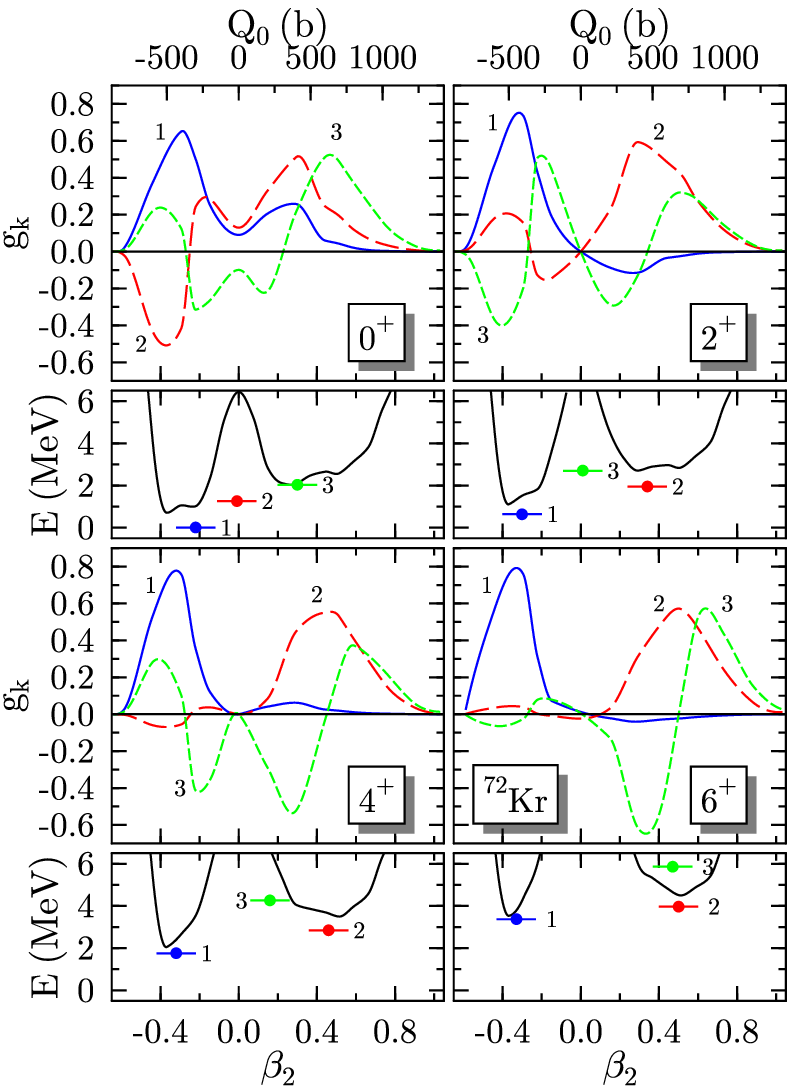}
\caption{\label{fig:kr72:wf}
(color online).
The same as Fig.\ \ref{fig:kr74:wf} for \nuc{72}{Kr}.
}
\end{figure}

\begin{figure}[t!]
\includegraphics{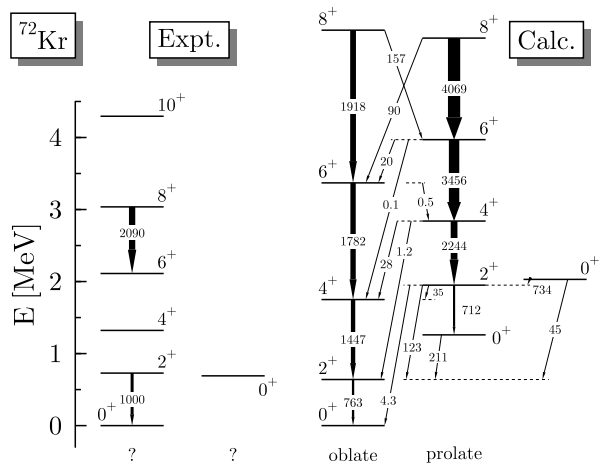}
\caption{\label{fig:kr72:collspec}
The same as Fig.\ \ref{fig:kr74:collspec} for \nuc{72}{Kr}. There is no
unambiguous assignment of the experimental yrast states into a prolate or
oblate band (see text).
Experimental data for the $B(E2)$ values are taken from \cite{Ang97aE}
($8^+ \to 6^+$) and \cite{Gad05aE} ($2^+_1 \to 0^+_1$).
}
\end{figure}

The potential energy curves, together with the GCM spectra, are shown 
in Fig.\ \ref{fig:kr72:ex} and the collective wave functions in 
Fig.\ \ref{fig:kr72:wf}. According to the usual interpretation of
the low-lying $0^+$ states, our calculation reproduces their order
with a dominantly oblate ground state. In Fig.\ \ref{fig:kr72:collspec}, 
the experimental and theoretical
excitation energies and $B(E2)$ values are compared. According to
the usual interpretation of the two low-lying $0^+$ states, our
calculation reproduces their order with a dominantly oblate ground
state. The first excited $2^+$ is at about the same excitation
energy as the experimental one, while the first excited $0^+$ is
slightly too high. Our calculated oblate states are yrast up to
\mbox{$J=6$}. With the exception the low-lying $0^+$ states, which
strongly mix oblate, spherical and prolate states, the mixing
between the prolate and oblate configurations is less important than
for the heavier Kr isotopes. Within the accuracy that can be
expected from our model, our $B(E2)$ value for the $2^+_{\text{obl}}
\to 0^+_{\text{obl}}$ transition is consistent with the experimental
one, $B(E2; 2^+_1 \to 0^+_1) = 1000 (130)$ $e^2$ fm$^4$.

\nuc{72}{Kr} is the only Kr isotope studied here for which
we can assume to reproduce the ordering of the low-lying states.
%
%
\subsection{$E0$ transitions and radii}
The nuclear matrix element entering the monopole decay rate of a
$J^\pi$ state to a state with the same spin and parity is given by
\cite{Chu56a}
\begin{equation}
\rho_{E0}^2 (J_{k'} \to J_k)
= \left| \frac{\langle JM k | \hat{r}^2 | J M k'\rangle}{R^2} \right|^2
,
\end{equation}
where \mbox{$R = 1.2 A^{1/3}$} fm. Within a simple
model~\cite{Hey88a}, this matrix element can be related to the
amount of mixing of configurations with different deformations in
the physical states.  As it is a non-diagonal matrix element, it is
also very sensitive to the detailed structure of the initial and
final states. Within the error bars, the experimental $\rho_{E0}^2$
values are very close for \nuc{72-76}{Kr}, and slightly smaller for
\nuc{78}{Kr}. The variation of the GCM values is much larger, but in
view of their sensitivity to model details it is encouraging that
our calculated $\rho_{E0}^2$ values are within the experimental
error bars for \nuc{74}{Kr} and \nuc{78}{Kr} and underestimate the
value for \nuc{72}{Kr} by only a factor two. The large
underestimation for \nuc{76}{Kr} can be related to the much smaller
mixing between the oblate and prolate $0^+$ states that we obtain
for this isotope.

\begin{figure}[t!]
\includegraphics{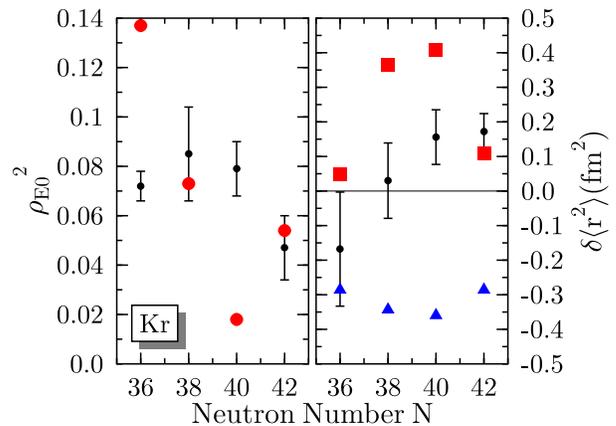}
\caption{\label{fig:all:e0}
(color online).
Left panel: comparison of calculated values for
$\rho_{E0}^2 (0^+_2 \to 0^+_1)$ with experiment. Experimental values are
taken from \cite{Bou03aE} (\nuc{72-74}{Kr}) and \cite{Gia95aE}
(\nuc{76-78}{Kr}). Right panel: isotopic
shift of the mean-square charge radii for the lowest (dominantly) oblate
$0^+$ state (triangles) and the lowest (dominantly) prolate $0^+$ state
(squares). The ground state predicted by our calculation is always oblate.
}
\end{figure}

The reasonable description of the $\rho_{E0}^2$ states indicates
that, with the exception of \nuc{76}{Kr}, we have about the right
amount of mixing between the oblate and prolate $0^+$ states, which
is independent of the relative placement of the energy levels. In
contrast, the systematics of ground-state radii is not well
described, as our calculations give the "wrong" ground state for
\nuc{74-78}{Kr}. This can be seen from the right panel of Fig.\
\ref{fig:all:e0}, where the isotopic shifts of the mean-square
charge radii
\begin{equation}
\delta r^2 (N) = r^2 (N) - r^2 (N=50),
\end{equation}
are plotted.
 The Kr radii present an anomaly  \cite{Kei95aE,Lie96aE}: they are
larger for \nuc{74-84}{Kr} than for the \mbox{$N=50$} isotope
\nuc{86}{Kr}. The standard interpretation is that the ground states
have an increasing admixture of prolate states of large deformation
when going to more neutron-deficient nuclei, which overcompensates
the volume effect of decreasing mass. As our calculations always
predict dominantly oblate ground states, we obtain a negative
isotopic shift for all for isotopes (triangles in the right panel of
Fig.\ \ref{fig:all:e0}). Our calculated excited dominantly prolate
$0^+$ states, by contrast, lead to positive isotopic shifts for all
four isotopes. It is unclear how well the ground state of
\nuc{86}{Kr} itself is described within our method, so one has to be
careful with the interpretation of the absolute values of the
isotopic shifts. However, the $\delta r^2$ data confirm once again
that the prolate states should have a much larger contribution to
the ground state.

The relativistic mean-field calculations with the NL-SH
interaction presented in Ref.\ \cite{Lal95a} do reproduce the
systematics of the isotopic shifts, which hints at the more
realistic single-particle spectrum of this interaction in the $fp$
shell.
%
%
\section{Discussion and Conclusions}
Our results show that in medium-mass nuclei with coexisting
shallow minima in soft potential energy landscapes, one has to be
very cautious when comparing experimental data and the mean-field
states corresponding to local minima. Projection on good angular
momentum alters significantly the potential energy surfaces. After
configuration mixing, the collective wave functions present a very
large spreading and may extend over several mean-field minima. We
also find rotational bands, for example in \nuc{78}{Kr}, which are
not related to a minimum in the mean-field energy landscape, but
to a plateau.

While our calculations reproduce many of the global features of the
neutron-deficient Kr isotopes, there remain several discrepancies
with experiment when looking in detail.

\begin{itemize}
\item
The collective spectra are too spread out in excitation energy,
which appears to be a general problem of GCM of angular-momentum
projected axial mean-field states \cite{Ben05b,Egi04a}. This problem
could probably be cured by a larger variational space for the
projected GCM, in particular by treating dynamically the pairing, by
breaking axial symmetry or time-reversal invariance. Breaking axial
symmetry will also permit the description of $\gamma$ bands.
Work in these various directions is underway.
\item
Our study of neutron-deficient Kr isotopes clearly points at a
problem with the SLy effective interactions in the $fp$ shell,
related to the relative position of a few single-particle levels.
\end{itemize}

As already discussed in the literature, (see, e.g., the discussion
in \cite{Lan03a}), the spectroscopic properties of  the
neutron-deficient Kr region are extremely sensitive to the details
of the shell structure. The number of levels around the Fermi energy
is quite small which results in several shell gaps for every neutron
and proton number from 34 to 42. Any change in the relative position
of levels and of their ordering will modify the size and deformation
of the gaps in Fig.\ \ref{fig:kr74:spe}.

Our results, and those of models with adjustable single-particle
spectra that perform better for Kr isotopes, suggest a few
necessary modifications of the single-particle spectra obtained
with the SLy6 Skyrme parametrization:
\begin{itemize}
\item
The experimental evidence that there are no spherical $0^+$ states
in the spectrum of \nuc{76}{Kr} shows that the \mbox{$N=40$}
spherical shell gap is too strong. This is consistent with our
results for the \mbox{$N \approx Z \approx 40$} region obtained in a
systematic study of mass and deformation where the ground state for
\nuc{80}{Zr} was obtained spherical with a very pronounced shell
effect \cite{BBH05a}, while experiment suggests that is a
well-deformed rotor \cite{Lis87aE}. A $g_{9/2}$ level closer to the
$p_{1/2}$ orbital would decrease the \mbox{$N=40$} spherical gap. It
would also decrease the oblate \mbox{$N=38$} gap, and shift the
prolate \mbox{$N=38$} gap to smaller deformation, probably reducing
the deformation of \nuc{74}{Kr}.

\item
Taking the Wodds-Saxon single-particle level schemes of \ Fig.\ 17 in
\cite{Mol97a} as an example, a decrease of the separation between
the $f_{5/2}$ and the $p_{3/2}$ levels should change the
single-particle spectra at small prolate deformation: a gap at
\mbox{$N=38$} that extends from oblate shapes to prolate shapes
 would replace the gap at \mbox{$N=36$} around $\beta_2=0.2$. It can
be expected that this would drive \nuc{74}{Kr} towards prolate
shapes.
\end{itemize}
We have checked that all modern successful Skyrme interactions give
single-particle spectra in the $fp$ shell which are similar to those
obtained with SLy6. This points to a deficiency of the standard
Skyrme interaction in general and not to a difficulty intrinsic to
SLy6. Other models as the relativistic mean-field model (RMF) might
seem to work more satisfactorily in the $fp$ shell and to predict
potential landscapes in better qualitative agreement with the Kr
data than ours \cite{Lal95a}. However, those models do not describe
correctly the neutron-deficient Pb region \cite{Hey96a,Nik02a},
where many Skyrme forces perform quite well \cite{Dug03pb,Ben03pb}.

Altogether, the present study confirms the conclusions drawn in
\cite{BBH05a} that the current functional form of the Skyrme
energy functional is not yet flexible enough to cover the relevant
aspects of nuclear structure with the same good quality for all
regions of the nuclear chart.

The combined analysis of Figs.\ \ref{fig:kr74:spe} and
\ref{fig:all:pes} suggests that the structure of collective states
in the neutron-deficient Krypton isotopes provides a sensitive
testing ground for future attempts to construct better effective
interactions. It has been pointed out recently that a tensor
interaction, absent in the existing standard mean-field models,
introduces a particle-number dependence of single-particle
spectra \cite{Otsukatalk,Brinktalk,Dobaczewskitalk}, that might
resolve at least some of the deficiencies in the $fp$ shell we
encounter here, and that are known for other regions of the
nuclear chart. Work in that direction is underway.

%
%
\begin{acknowledgments}
We thank  E.\ Bouchez, E. Cl{\'e}ment, A.\ Gade, A.\ G{\"o}rgen and
W.\ Korten for stimulating discussions on the available experimental
data and for making their results available to us prior to
publication. We also thank R.\ V.\ F.\ Janssens and C.J.\ Lister for
interesting discussions. This work was supported by the Belgian
Science Policy Office under contract PAI P5-07 and the US National
Science Foundation under Grant No.\ PHY--0456903. MB acknowledges
financial support from the European Commission. MB and PHH thank for
the hospitality at the Institute for Nuclear Theory, Seattle, where
part of this work was carried out.
\end{acknowledgments}
%
%

\end{document}